\documentstyle[prl,aps,preprint]{revtex}
\begin{document}
\draft
\title{{\bf 
Flux-line entanglement as the mechanism of melting transition \\
in high-temperature superconductors in a magnetic field
}}
\author{Yoshihiko Nonomura, Xiao Hu, and Masashi Tachiki}
\address{
National Research Institute for Metals, Tsukuba, Ibaraki 305-0047, Japan
}
\date{Received: 06 Nov.\ 1998, revised 10 Feb.\ 1999, Phys.\ Rev.\ {\bf B} 
Rapid Commun.\ in press}
\maketitle
\begin{abstract}
The mechanism of the flux-line-lattice (FLL) melting in anisotropic 
high-$T_{{\rm c}}$ superconductors in ${\bf B}\parallel {\bf \hat{c}}$
is clarified by Monte Carlo simulations of the 3D frustrated XY model. 
The percentage of entangled flux lines abruptly changes at the melting 
temperature $T_{{\rm m}}$, while no sharp change can be found in the 
number and size distribution of vortex loops around $T_{{\rm m}}$. 
Therefore, the origin of this melting transition is the entanglement 
of flux lines. Scaling behaviors of physical quantities are consistent 
with the above mechanism of the FLL melting. The Lindemann number is 
also evaluated without any phenomenological arguments. 
\end{abstract}
\narrowtext
Nature of the mixed phase in type-II superconductors has been studied 
for many years, and much attention has been paid to this field since 
the discovery of high-$T_{{\rm c}}$ superconductors (HTSC) because of 
short correlation lengths and large anisotropy. In HTSC in a magnetic 
field along the $c$ axis, the flux-line lattice (FLL) melts at much 
lower temperatures or in much weaker fields \cite{HTSCrev} than 
those predicted by the Abrikosov mean-field theory, where the 
superconducting phase transition is of second order regardless of 
details of models. The FLL melting in HTSC was first theoretically 
analyzed by Nelson and Seung \cite{Nelson1} on the basis of the 
mapping to a two-dimensional Boson model, and they pointed out 
that the Lindemann criterion might be valid in this FLL melting 
because of large fluctuations in HTSC owing to large anisotropy 
and high transition temperatures. Similar theoretical analysis was 
also made by Houghton {\it et al.} \cite{Houghton} independently. 
Earlier experiments of ``the FLL melting in HTSC" as reviewed 
in Ref.\ 1 were found to be explained better by the vortex-glass 
transition \cite{Fishers} rather than by the FLL melting 
transition. In clean systems, the ``true" FLL melting was 
confirmed afterwards by experiments 
\cite{Safar,Kwok,Welp,Fendrich,Zeldov,Schilling} 
and computer simulations 
\cite{Hetzel,Sasik,Ryu1,Ryu2,Hu,HuJ,Nordborg,Koshelev1}, 
and the first-order FLL melting transition in a magnetic field has now 
been established in extremely type-II superconductors such as HTSC. 

On the other hand, the mechanism of the FLL melting has not yet been well 
understood. Nelson argued \cite{Nelson1,Nelson2} that the entanglement 
of flux lines is related to this transition, but this picture has only been 
confirmed numerically \cite{Nordborg} within the two-dimensional Boson 
model. Thermal excitations of vortex loops \cite{Tesanovic1,Nguyen1} 
are not included in this picture and some authors claimed 
\cite{Nguyen1,Nguyen2} that the FLL melting is characterized by the 
breakdown of flux-line description by the proliferation of large vortex loops, 
but the latter picture began to be modified recently \cite{Tesanovic2}. 
In order to clarify the mechanism of the FLL melting numerically, 
the first-order phase transition should be identified by measuring 
thermodynamic quantities, and the behavior of flux lines induced by 
an external magnetic field and thermally excited vortex loops should 
be observed microscopically in the vicinity of the melting temperature. 

In this article, the three-dimensional anisotropic, frustrated 
XY model is analyzed with the Monte Carlo method from 
the above point of view. Our main results are as follows: 
First, the mechanism of the first-order FLL melting 
transition is exclusively the entanglement of flux lines. 
The percentage of entangled flux lines sharply changes at 
the melting temperature $T_{{\rm m}}$, while the number and 
the size distribution of loop excitations has a smooth temperature 
dependence around $T_{{\rm m}}$. Second, scaling properties around 
the melting temperature are clarified. As a consequence of the 
entanglement mechanism of the FLL melting, $T_{{\rm m}}$ 
is scaled by the inverse of the system size along the $c$ axis. 
Third, the Lindemann number takes nearly a constant value 
$c_{{\rm L}}\approx 0.30$ regardless of the anisotropy constant, 
and thus the use of the Lindemann criterion is justified for the 
determination of the melting line in a phase diagram. 

As the model of the anisotropic, extremely type-II HTSC in a magnetic 
field along the $c$ axis, we consider the three-dimensional anisotropic, 
frustrated XY model \cite{Li,Hu} described by the following Hamiltonian, 
\begin{eqnarray}
{\cal H}&=&-J \hspace{-0.2cm}
                 \sum_{i,j \in ab\ {\rm plane}} \hspace{-0.3cm}
                 \cos \left(\varphi_{i}-\varphi_{j}-A_{ij}\right)
                 \nonumber\\
           &&-\frac{J}{\Gamma^2} \hspace{-0.1cm}
               \sum_{i,j \parallel c\ {\rm axis}}\hspace{-0.2cm}
               \cos \left(\varphi_{i}-\varphi_{j}\right)\ ,\\
A_{ij}&=&\frac{2\pi}{\phi_0}\int^{j}_{i}{\bf A}^{(2)}
                                      \cdot {\rm d}{\bf r}^{(2)}.
\end{eqnarray}
Here $\varphi_{i}$ denotes the phase of the superconducting 
order parameter, $\phi_{0}$ stands for the flux quantum, and 
the anisotropy is represented by the parameter $1/\Gamma^{2}$. 
Note that this model neglects fluctuations of the gauge field 
and the amplitude of the superconducting order parameter. 
The periodic boundary condition (PBC) is applied on the 
phase variable $\varphi_{i}$ in all the directions in order 
to refrain from finite-size effects from free boundaries. 
We mainly investigate the case with the averaged number 
of fluxes per plaquette $f=1/25$, $\Gamma=2$ and $5$. 
In these parameters, effects of the introduction of a square 
lattice in the $ab$ plane is negligible \cite{Hu}. Each phase 
variable takes a value $-\pi<\varphi_{i} \leq \pi$, and the 
summation of the phase difference around a plaquette is given by 
\begin{equation}
  \sum_{i,j\in \Box}\left(\varphi_{i}-\varphi_{j}-A_{ij}\right)
 =2\pi(n-f)\ ,\ \ n=1,\ 0,\ -1\ .
\end{equation}
When the integer $n$ takes $1$ or $-1$, the plaquette is defined to 
have a vortex or an antivortex, respectively. The nearest-neighbor 
vortices are connected with one another to form vortex lines, 
which do not have end points inside the system. When a vortex 
line returns to itself inside the system, it is called as a 
{\it vortex loop}, and such a loop does not exist in the ground 
state. When a vortex line runs from one boundary to another along 
the direction of the external field, it is called as a {\it flux line}. 
The flux lines are straight in the ground state, and they begin 
to fluctuate at finite temperatures. A flux line is defined as 
{\it entangled} if it does not terminate at the same transverse 
position in the top and bottom boundaries in the PBC. 
The flux lines which wind with each other and return to the initial 
transverse positions inside the system are not included in this 
definition. Such excitations are negligible in the vicinity of the 
melting temperature, because they have higher energies. 

The helicity modulus along the $c$ axis \cite{Hu,Michel} is observed 
for the determination of the melting temperature. This quantity is 
proportional to the superfluid density, and therefore nonvanishing 
only in the superconducting phase. The numbers of vortex loops 
$N_{{\rm loop}}$ and entangled flux lines $N_{{\rm ent}}$ 
are counted. The distribution of sizes is also measured for 
Josephson loops, which are dominant below and slightly above 
the melting temperature. The transverse distance $w$ of a 
flux line between the top and bottom $ab$ planes is measured, 
and its averaged value over all the flux lines is denoted by 
$L_{{\rm diff}}$. The fluctuation of a flux line is measured 
by the deviation $u$ from the projection of its mass center in 
each $ab$ plane, and averaged over all the flux lines and the 
$ab$ planes. The Lindemann number $c_{{\rm L}}$ is defined by 
\begin{equation}
  \label{Lindef}
  c_{{\rm L}}\equiv\lim_{T\to T_{{\rm m}}-0}
                   \langle u^{2}\rangle^{1/2}/a_{0}\ ,
\end{equation}
where $a_{0}$ stands for the lattice constant of the triangular FLL, 
$a_{0}=(2/\sqrt{3})^{1/2}/f^{1/2}$. 

Monte Carlo simulations are performed on the basis of the Metropolis 
algorithm. Most results reported in this article are for systems with 
$L_{x}=L_{y}=50$ and $L_{c}=80$. Then, the number of total flux lines 
is $N_{{\rm flux}}=100$ in the present simulations with $f=1/25$. 
In order to check the size dependence, we also simulate systems 
with $(L_{x},L_{c})=(50,20)$, $(50,40)$, $(50,54)$, $(50,160)$, 
$(25,80)$  and $(100,20)$. In addition, in order to check the 
flux-density dependence, the $(L_{x},L_{c})=(50,80)$ and 
$(100,80)$ systems are calculated for $f=1/50$ and $1/100$, 
respectively. Simulations are started from temperatures 
more than ten times higher than $T_{{\rm m}}$ and the system 
is gradually cooled down. Typical Monte Carlo steps (MCS) are 
$1.0\times 10^{5}$ and $1.5\times 10^{5}$ for equilibration (E-MCS) 
and measurement (M-MCS) at each temperature, respectively. 
Since the correlation time becomes longer in the vicinity of the 
melting point, E-MCS and M-MCS are taken as $3.5\times 10^{5}$ 
and $8.0\times 10^{5}$, respectively, and the cooling rate 
is reduced to $\Delta T=1.0\times 10^{-3} J/k_{{\rm B}}$. 
Moreover, for the precise determination of $T_{{\rm m}}$ and 
$c_{{\rm L}}$, supercooling behavior is reduced very carefully 
by using up to $1.0\times 10^{7}$ MCS at each temperature. 
The helicity modulus is measured at each MCS, and the numbers of 
vortex loops and entangled flux lines are measured once per 100 MCS. 

The temperature dependence of the helicity modulus along the $c$ axis, 
$\Upsilon_{c}$, is displayed in Fig.~\ref{totfig} for $\Gamma=2$ and $5$ 
with $f=1/25$. This quantity sharply drops from a finite value to zero 
at the melting temperature, $T_{{\rm m}}\simeq 0.810 J/k_{{\rm B}}$ for 
$\Gamma=2$ and $T_{{\rm m}}\simeq 0.3445 J/k_{{\rm B}}$ for $\Gamma=5$, 
which indicates the thermodynamic first-order phase transition \cite{Hu}. 
The temperature dependence of the ratio of entangled flux lines to 
total flux lines, $N_{{\rm ent}}/N_{{\rm flux}}$, is also displayed 
in Fig.~\ref{totfig}. It shows a sharp jump at $T_{{\rm m}}$ for each 
of the anisotropy. Similar behavior is also observed for $\Gamma=5$ 
with $f=1/50$ and $1/100$. The number of vortex loops per flux line 
per $ab$ plane, $N_{{\rm loop}}/(N_{{\rm flux}}L_{c})$, is shown in 
Fig.~\ref{loopfig} for $\Gamma=2$ and $5$ with $f=1/25$. The temperature 
dependence of this quantity is not as drastic as that of the ratio of 
entangled flux lines. The size distribution of Josephson loops is 
also measured for $\Gamma=5$ with $f=1/25$, $1/50$ and $1/100$, 
and no drastic change is observed in this distribution around 
$T_{{\rm m}}$. The numbers of vortex loops are not of the same order 
for the different anisotropy constants at the melting temperatures. 
That is, the number of vortex loops at $T_{{\rm m}}$ for $\Gamma=5$ 
corresponds to that of $T\approx 0.6 J/k_{{\rm B}}$ for $\Gamma=2$, 
and this temperature is much lower than the melting point for 
$\Gamma=2$, $T_{{\rm m}}\simeq 0.810 J/k_{{\rm B}}$. 
These facts clearly indicate that the origin of the FLL melting 
is the entanglement of flux lines, at least up to $f=1/100$. 

Then, we show the results for the finite-size-scaling behavior. 
The averaged end-to-end transverse distance of flux lines, 
$L_{{\rm diff}}$, is normalized by the lattice constant of FLL, 
$a_{0}$, and plotted versus temperature for $L_{c}=40$, $80$ 
and $160$ in Fig.~\ref{sizefig}. The size dependence of this 
quantity can be described by the random-walk-type scaling, 
\begin{equation}
  \label{RWsc}
  L_{{\rm diff}}(L_{c})\sim {\rm const.}\times L_{c}^{1/2},
\end{equation}
for a certain temperature range above $T_{{\rm m}}$, as 
displayed in the inset of Fig.~\ref{sizefig}. Therefore, it is clear 
that the vortices form flux lines even above $T_{{\rm m}}$. 
The inset of Fig.~\ref{sizefig} shows that the data for 
$L_{c}=160$ and $80$ deviate from the scaling (\ref{RWsc}) 
at $T\approx 0.4 J/k_{{\rm B}}$ and $0.5 J/k_{{\rm B}}$, 
respectively. These two temperatures correspond to the same 
transverse distance $L_{{\rm diff}}\approx 2.7 a_{0}$, as 
can be read from the two curves for $L_{c}=160$ and $80$ 
in the main body of Fig.~\ref{sizefig}. This fact suggests 
that the random-walk behavior is restricted within a transverse 
diffusion distance approximately $2.7 a_{0}$ independently of 
temperature even in bulk systems. Beyond this length scale, 
reconnection between flux lines occurs frequently, and the 
random-walk property is suppressed. Since the transverse 
distance between the bottom and top $ab$ planes in each 
entangled flux line cannot be smaller than $a_{0}$, the total 
number of entangled flux lines abruptly decreases at $T_{{\rm m}}$ 
when temperature is gradually reduced. As a consequence of the 
scaling given in Eq.~(\ref{RWsc}), the temperature characterized 
by $L_{{\rm diff}}\approx a_{0}$, namely $T_{{\rm m}}$, 
depends on $L_{c}$, as will be discussed below. 

Simulations for systems with $(L_{x},L_{c})=(100,20)$ and 
$(25,80)$ are also performed, and the results coincide with 
those with $(L_{x},L_{c})=(50,20)$ and $(50,80)$, respectively. 
This is quite natural because the melting transition is 
characterized by the entanglement of flux lines along the $c$ 
axis, and the leading term of size dependence is only related 
to $L_{c}$. Finite-size effects in the $ab$ plane are indirect 
on thermodynamic quantities. As displayed in Fig.~\ref{FSSfig}, 
our data exhibit the following finite-size scaling, 
\begin{equation}
  \delta T_{{\rm m}}(L_{c})
  \equiv T_{{\rm m}}(L_{c})-T_{{\rm m}}(L_{c}=\infty)
  \sim {\rm const.}\times L_{c}^{-1}, 
\end{equation}
with $T_{{\rm m}}(L_{c}=\infty)=0.3354\pm 0.0007$ for 
$\Gamma=5$. This size dependence means that the FLL melting 
has a one-dimensional character, because the scaling form of the 
transition point in first-order phase transitions \cite{1dfss} is 
generally given by $\delta T(L)\sim {\rm const.}\times L^{-D}$ 
with the spatial dimension $D$. This one-dimensional character 
is consistent with the entanglement picture. 

Finally, we turn to see the temperature dependence of the 
fluctuation of flux lines and evaluate the Lindemann number. 
The quantity $\langle u^{2}\rangle^{1/2}/a_{0}$ shows 
a sharp jump at $T_{{\rm m}}$ for $\Gamma=2$ and $5$ 
with $f=1/25$ (Fig.~\ref{Linfig}). From the definition 
of the Lindemann number in Eq.~(\ref{Lindef}), we have 
$c_{{\rm L}}\approx 0.30$ for both the anisotropy constants. This 
result suggests that the Lindemann number does not depend on details 
of models, as assumed in previous studies \cite{Houghton,Blatter}. 
The Lindemann number was evaluated as $c_{{\rm L}}\approx 0.18$ 
by part of the present authors \cite{Hu} by fitting the simulated 
melting line with a formula derived by Blatter {\it et al.} 
\cite{Blatter} based on the London theory. We believe that the 
present direct evaluation of $c_{{\rm L}}$ is more reliable. 

In conclusion, the three-dimensional anisotropic, frustrated XY model 
has been analyzed with Monte Carlo simulations. The melting 
temperature $T_{{\rm m}}$ has been estimated as the point 
at which the helicity modulus along the $c$ axis vanishes. 
The percentage of entangled flux lines shows a sharp jump 
at $T_{{\rm m}}$, while the number and size distribution of 
vortex loops do not show such drastic change at $T_{{\rm m}}$. 
This fact clearly indicates that the origin of the FLL melting in a 
magnetic field along the $c$ axis is the entanglement of flux lines. 
The consistency of this picture with the size dependence of 
various quantities has been confirmed. Especially, the melting 
temperature is scaled by the system size along the $c$ axis as 
$T_{{\rm m}}(L_{c})-T_{{\rm m}}(L_{c}=\infty)\propto L_{c}^{-1}$. 
The averaged deviation of flux lines from their mass centers also 
shows a sharp jump at $T_{{\rm m}}$ as a consequence of the 
entanglement of flux lines. The Lindemann number takes a constant 
value $c_{{\rm L}}\approx 0.30$ regardless of the anisotropy. 
This numerical result justifies the use of the Lindemann 
criterion for characterizing the FLL melting in HTSC. 

Numerical calculations were performed on the Numerical Materials 
Simulator (NEC SX-4) at National Research Institute for Metals, Japan. 
\noindent
\begin{figure}
\caption{
Helicity modulus along the $c$ axis (squares) and the ratio 
of entangled flux lines to total flux lines (circles) versus 
temperature for $\Gamma=2$ and $5$ with $f=1/25$. 
}
\label{totfig}
\end{figure}
\begin{figure}
\caption{
Normalized number of vortex loops (diamonds) versus temperature 
for $\Gamma=2$ and $5$ with $f=1/25$. The ratio of entangled 
flux lines (circles) is also plotted for comparison. 
}
\label{loopfig}
\end{figure}
\begin{figure}
\caption{
Normalized end-to-end transverse distance versus temperature 
for $L_{c}=40$ (triangles), $80$ (squares) and $160$ (circles) for 
$\Gamma=5$ with $f=1/25$. Scaling plot of the same data according 
to Eq.\ (\ref{RWsc}) is shown in the inset with the same symbols. 
}
\label{sizefig}
\end{figure}
\begin{figure}
\caption{
$L_{c}$ dependence of the melting temperature for $\Gamma=5$ with 
$f=1/25$. $T_{{\rm m}}(L_{c}=\infty)$ is estimated by the least-squares 
fitting of the data with $L_{c}=40$, $54$, $80$ and $160$. 
}
\label{FSSfig}
\end{figure}
\begin{figure}
\caption{
Fluctuation of flux lines (triangles) versus temperature for 
$\Gamma=2$ and $5$ with $f=1/25$. The helicity modulus 
along the $c$ axis (squares) is also plotted for comparison. 
}
\label{Linfig}
\end{figure}
\end{document}